%% file: main.tex
\documentclass[reprint,superscriptaddress,amsmath,amssymb,aps,prl]{revtex4-2}

\usepackage{graphicx}
\usepackage{dcolumn}
\usepackage{bm}
\usepackage{physics}
\usepackage{xcolor}

\usepackage{hyperref} 
\hypersetup{colorlinks, allcolors=black}

\makeatletter
\newcommand\footnoteref[1]{\protected@xdef\@thefnmark{\ref{#1}}\@footnotemark}
\makeatother

\begin{document}

\preprint{APS/123-QED}

\title{Spectroscopy and topological properties of a Haldane light system}

\author{Julian Legendre}
%\email{julian.legendre@uni.lu}
\affiliation{Department of Physics and Materials Science, University of Luxembourg, L-1511, Luxembourg}
\affiliation{CPHT, CNRS, Institut Polytechnique de Paris, Route de Saclay, 91128 Palaiseau, France}
\author{Karyn Le Hur}
 %\email{karyn.le-hur@polytechnique.edu}
\affiliation{CPHT, CNRS, Institut Polytechnique de Paris, Route de Saclay, 91128 Palaiseau, France}

%\date{\today}

\begin{abstract}

We introduce a local spectroscopic method in real space to probe the topological properties of a circuit quantum electrodynamics (cQED) array generalizing previous approaches from one to two dimensions in the plane. As an application, we develop the theory of microwave light propagating in the local probe capacitively coupled to the cQED array associated to a bosonic Haldane model. Interestingly, we show that the measured reflection coefficient, resolved in frequency through the resonance, reveals the model’s geometrical properties and topological phase transition. We discuss the role of physical parameters such as the lifetime of the light modes and stability towards local disorder related to further realizations. 
\end{abstract}

\maketitle

\paragraph*{Introduction.---} Topological systems find various interesting applications in physics, in particular related to the protected mesoscopic transport at the edges. In two dimensions, the quantum Hall effect, induced by a perpendicular uniform magnetic field, has been generalized to situations with no net flux in a unit cell, referring to the Haldane honeycomb lattice model \cite{Haldane88}, and then generally to the quantum anomalous Hall effect and Chern insulators. The latter are realized in solid-state systems, in cold atom gases and in photonic systems (coupled waveguides) \cite{Chang13,Rechtsman13,Jotzu14}. One elegant way to realize Haldane's seminal model for artificial systems is through Floquet engineering \cite{Takashi09,Rechtsman13,Jotzu14,Zheng14,Eckardt15,Plekhanov17}.

The most common way to probe the topological properties in condensed matter systems is to determine the Hall conductance \cite{Klitzing80,Chang22}. 
The topological responses of artificial systems are accessible in several ways \cite{Goldman16,Cooper19,LuJ14,OzawaPrice19}.
In cold atom gases, topological properties are revealed through transport or Hall drift \cite{Jotzu14,Aidelsburger15}, interferometry
\cite{Atala13,Duca15,TracyLi16}, the physics of chiral edge states \cite{Stuhl15,Mancini15} or via a measurement of the Berry curvature \cite{Flaschner16}.
For condensed matter systems and cold atom gases, a circular drive on the system also enables to probe the topological information \cite{Mciver12,Juan17,Tran17,Asteria19,Rees20}, even with a local resolution within the Brillouin zone \cite{KleinP21,HurK22}.
Light systems with topological properties, including gyromagnetic photonic crystals \cite{Haldane08,Raghu08,WangChong08,WangChong09}, arrays of coupled waveguides \cite{Rechtsman13,Mukherjee17,Maczewsky17}, optomechanical systems \cite{Fleury14,Kim17}, cavity and circuit quantum electrodynamics (cQED) \cite{Koch10,Anderson16,Owens18,Hur16,GuKockum17,Roushan17}, have also garnered significant interest. 

In Ref.~\cite{Goren18}, a protocol to probe the topological properties of a one-dimensional LC circuit system is proposed. This system is closely connected to the SSH model which has been implemented recently \cite{Rosenthal18,Poli15,Meier16,StJean17}. In Ref.~\cite{Goren18}, the authors considered a transmission line (capacitively) coupled to a single cell within the chain. From the reflection of an input triggered in the probe, they reconstructed the Zak phase, which is the topological invariant characterizing the studied one-dimensional system. 
Our quest is to generalize this local probe approach on the lattice in two dimensions, which is {\it \` a priori} not so apparent. Previous proposals for light-matter topological probes in two-dimensional systems have used the transverse polarization of light to detect the chirality associated with the system's topological nature \cite{Mciver12,Tran17}. In striking contrast to these approaches, our study focuses on a local probe in real space, specifically a long transmission line capacitively coupled to a Haldane bosonic model in circuit quantum electrodynamics (cQED). We demonstrate how the Chern number can be measured by analyzing the reflection coefficient, which relates the input and output voltage signals.

\begin{figure}[b]
\includegraphics[width=\linewidth]{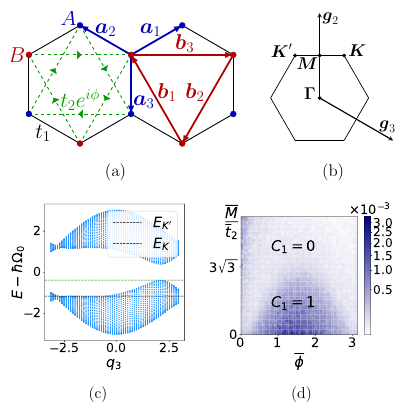}
\caption{
\label{fig1} (a) Definition of the sublattices, real space vectors and hopping amplitudes for the Haldane model on the honeycomb lattice. (b) Sketch of the Brillouin zone for the honeycomb lattice. (c) Haldane model energies, in units of $t_1$, as a function of the momentum $q_3 = a {\bf k} \cdot {\bf g}_3$ ($a$ is the lattice spacing), with parameters $\phi=\pi/2$, $t_2 = 0.15 t_1$ and $M=3 \sqrt{3}t_2/2$. Both lowest band energies at the Dirac points, $E_{1,{\bf K}}$ and $E_{1,{\bf K}'}$, are shown. (d) The quantity $1-\int d \omega S^{\textrm{out}}(\omega)$, computed for a disordered $21 \times 20$ unit cells Haldane system with open boundary conditions. The color scale of the figure is logarithmic. The precise information on definitions and parameters can be found page 4.}
\end{figure}

\paragraph*{Bosonic Haldane model.---} We introduce a cQED system made of an array of resonators coupled together in such a way \cite{Plekhanov17} that the system is described by a usual Haldane Hamiltonian $H= \sum_{{\bf k}} \Psi_{{\bf k}}^\dagger h_{{\bf k}} \Psi_{{\bf k}}$ \cite{Haldane88,Chang13,Rechtsman13,Jotzu14,Takashi09,Zheng14,Eckardt15}, with
\begin{equation} \label{eq:hk}
h_{{\bf k}} = h_0({\bf k})   + \textrm{Re}\left[h_1({\bf k}) \right]\sigma_x -\textrm{Im}\left[h_1({\bf k}) \right] \sigma_y + h_2({\bf k})  \sigma_z,
\end{equation}
and $h_0({\bf k}) = \hbar \Omega_0  + 2 t_2 \cos \phi \sum_{i =1}^3 \cos ({\bf k} \cdot {\bf b}_i)$, $h_1({\bf k}) = t_1 \sum_{i =1}^3 \exp (-i {\bf k} \cdot {\bf a}_i)$, $h_2({\bf k}) = M- 2 t_2 \sin \phi \sum_{i =1}^3 \sin ({\bf k} \cdot {\bf b}_i)$ and $\Psi_{{\bf k}}^\dagger = \left(a_{1,{\bf k}}^\dagger, a_{2,{\bf k}}^\dagger\right)$, where $a^{\dagger}_{j,{\bf k}}$ is the creation operator for a boson with momentum ${\bf k}$ on sublattice $j$ ($j=1(2)$ corresponds to the sublattice A(B) appearing in Fig.~\ref{fig1}(a)). ${\bf a}_i$ and ${\bf b}_i$ ($i \in \{1,2,3\}$) are defined in Fig.~\ref{fig1}(a), the hopping amplitudes $t_1$ and $t_2$ and the Semenoff mass $M$ \cite{Semenoff84} are real numbers and $\sigma^{x},\sigma^{y},\sigma^{z}$ are Pauli matrices acting in sub-lattice space. 
Hereafter, we study the case where $t_2$ is small compared to $t_1$, as it is often the situation in physical systems. In Ref.~\cite{Plekhanov17}, a Haldane Hamiltonian for bosonic systems is derived from Floquet engineering with a high-frequency approximation.
For a photonic system, a permanent drive is necessary to compensate for the photon decay processes that happen \cite{OzawaPrice19}. In typical photonic systems
the on-site energy $\hbar\Omega_0$ is large (usually $\sim$ GHz order of magnitude) compared to the effective hopping amplitudes on the lattice (\textit{e.g.} can be $\sim 10$ MHz to $\sim 100$ MHz) \cite{Koch10,Underwood12,Hartmann16,Roushan17}.

The Haldane model shows two energy bands in momentum space $E_{i,{\bf k}} = h_0({\bf k}) + (-1)^i  \epsilon({\bf k})$, where $i=1$ or 2 and $\epsilon({\bf k}) = \sqrt{|h_1({\bf k})|^2+h_2({\bf k})^2}$ (see Fig.~\ref{fig1}(c)).
%Band crossing appears at $h_1({\bf k}) = h_2({\bf k}) = 0$. 
$h_1({\bf k}) =0$ is reached at both nonequivalent Dirac points ${\bf K} = \left({\bf g}_3-{\bf g}_2\right)/3$ and ${\bf K}'=\left({\bf g}_2-{\bf g}_3\right)/3$ (see Fig.~\ref{fig1}(b)). Moreover, we have $h_2({\bf K}) = 0$ if $M = + 3 \sqrt{3}  t_2 \sin \phi$ and $h_2({\bf K}') = 0$ if $M = - 3 \sqrt{3}  t_2  \sin \phi$. When the bands cross, the dispersion relation around ${\bf K}$ and ${\bf K}'$ is linear.

%%%%%%%%%%%%%%%%%%%%%%%%%%%%%%%%%%%%%%%%%%%%%%%%%%%%%%%%%%%
%%%%%%%%%%%%%%%%%%%%%%%%%%%%%%%%%%%%%%%%%%%%%%%%%%%%%%%%%%%

\paragraph*{ Topological properties.---} 

%%%%%%%%%%%%%%%%%%%%%%%%%%%%%%%%%%%%%%%%%%%%%%%%%%%%%%%%%%%%%%%%%%%%%%
\begin{table}[b]
\caption{\label{tab:sgnh2}Sign of $h_2$ at the Dirac points, as a function of $M$ and $\textrm{sgn}(\sin \phi)$.}
\begin{ruledtabular}
\begin{tabular}{ccc}
& $|M|<3 \sqrt{3} t_2 |\sin \phi|$ & $|M|>3 \sqrt{3} t_2 |\sin \phi|$ \\
\colrule
$\textrm{sgn} \, h_2({\bf K})$ & $-\textrm{sgn}(\sin \phi)$ 
&  $\textrm{sgn} \, M$\\
$\textrm{sgn} \, h_2({\bf K}')$ &  $\textrm{sgn}(\sin \phi)$  &  $\textrm{sgn} \, M$  \\
\end{tabular}
\end{ruledtabular}
\end{table}
%%%%%%%%%%%%%%%%%%%%%%%%%%%%%%%%%%%%%%%%%%%%%%%%%%%%%%%%%%%%%%%%%%%%

Here, we describe the geometrical properties of the system through Bloch eigenvectors and through a definition of the topological number resolved at the Dirac points in Eq. (\ref{eq:chnumber}). This definition is generally valid for a model that can be written as a 
spin-$\frac{1}{2}$ particle in momentum space \cite{KLHreview}. We introduce
$\ket{u_{i,{\bf k}}}$ as the Bloch eigenvectors of the Haldane Hamiltonian, \textit{i.e.} $\textrm{e}^{-i {\bf k} \cdot \hat{{\bf r}}} H   \textrm{e}^{i {\bf k} \cdot \hat{{\bf r}}} \ket{ u_{i,{\bf k}}} = E_{i,{\bf k}} \ket{ u_{i,{\bf k}}}$ and we define the coefficients $\alpha_{i }^{1}({\bf k})$ and $\alpha_{i }^{2}({\bf k})$ such that $
\ket{u_{i,{\bf k}}} = \left[\alpha_{i}^{1}({\bf k}) a_{1,{\bf k}}^{\dagger} + \alpha_{i}^{2}({\bf k}) a_{2,{\bf k}}^{\dagger} \right] \ket{0}
$. These coefficients may vanish only at the Dirac points. If the sign of $h_2$ is opposite at the non-equivalent Dirac points ($|M|< 3 \sqrt{3} t_2 |\sin \phi|$, see Table~\ref{tab:sgnh2}), then $\alpha_{i }^{1}({\bf k})$ vanish at one Dirac point and $\alpha_{i }^{2}({\bf k})$ vanish at the other Dirac point. It follows that it is impossible to find a unique and smooth phase  over all the BZ for the Bloch state $\ket{u_{i,{\bf k}}}$. This characteristic feature of Chern insulators \cite{Kohmoto85} forms the basis of the probe proposed in this letter, rendering it \textit{\` a priori} relevant for Chern insulators with non-degenerate Dirac points in the energy spectrum (for the Haldane system we consider this imposes $t_2 \lesssim 0.15t_1$). In this phase, the Chern number $C_i$ ($i$ is the band index) reads $C_{i} =(-1)^{i+1} \, \textrm{sgn}( \sin \phi)$. In the Supplemental Material, for completeness, we present a derivation of this formula \cite{suppl}. Now, we build an alternative definition of the topological number from the Dirac points. If the sign of $h_2$ is the same at both Dirac points ($|M|> 3 \sqrt{3} t_2 |\sin \phi|$, see Table~\ref{tab:sgnh2}), then $\alpha_{i }^{1}({\bf k})$ or $\alpha_{i }^{2}({\bf k})$ can be chosen non-zero over all the BZ. In this case, the Chern numbers $C_i$ are vanishing. From this analysis, for $|M| \neq 3 \sqrt{3} t_2 |\sin \phi|$, we have $C_i = (-1)^{i+1} \, \textrm{sgn}( \sin \phi ) \left[1- \textrm{sgn}( h_2({\bf K}) h_2({\bf K}')\right]/2$, \textit{i.e.} 
\begin{equation} \label{eq:chnumber}
    C_i = \dfrac{(-1)^{i}}{2}  \left[\textrm{sgn} \, h_2({\bf K}) - \textrm{sgn} \, h_2({\bf K}') \right].
\end{equation}
This formula has indeed a simple physical understanding for a Hamiltonian $h_{\bf k}$ written as a $2\times 2$ matrix. From the Ehrenfest theorem and a Bloch sphere correspondence the topological number is equivalent to
$C_i = (-1)^{i} \left[\langle \sigma_z(0)\rangle - \langle \sigma_z(\pi)\rangle\right]/2$ with $\langle \sigma_z\rangle=(-1)^i\cos\theta=(-1)^i\hbox{sgn} \, h_2(\theta)$ \cite{KLHreview}. In the following, we namely rely on Eq.~\eqref{eq:chnumber} to show how $C_i$ can be probed from the reflected light in a local probe capacitively coupled to a Haldane photonic system.
The simple idea behind our proposal is that the topological properties manifest as discernible sublattice weight variations of the wave function, enabling to reveal the topological transition through the coupling of a probe to one of the sublattice sites. 

We emphasize here that in Ref.~\cite{Goren18}, we proposed a capacitively coupled topological probe for a 1D system. It gives access to a phase whose winding around the BZ is the topological invariant (Zak phase). The probe proposed in the following is substantially different since it measures the information in Eq. (\ref{eq:chnumber}). 

\paragraph*{ Spectroscopic probe.---}
Here, we introduce the local spectroscopic approach, i.e. a local light probe with weak capacitive coupling to a Haldane boson system at position ${\bf R}_0$, on the sublattice $j_0$. 
The probe is a resonator with a certain number of (relevant) modes, described by the Hamiltonian $
H_{\textrm{prb}} =\sum_q \hbar \omega_q b_{q}^\dagger b_{q}$,
with $ b_q^{\dagger}, b_{q}$ the creation, annihilation operators for the mode $q$ characterized by the frequency $\omega_q$. The coupling is described by $H_{\textrm{cpl}} = \left( a_{{\bf R}_0}+ a_{{\bf R}_0}^\dagger \right)\sum_q g_q \left( b_{q}+ b_{q}^\dagger  \right)$, where $ a_{{\bf R}_0}^\dagger$ is the Fourier transform of $a^{\dagger}_{j,{\bf k}}$ at position ${\bf R}_0$. For simplicity, we initially disregard the dissipation effects induced by the probe in the Haldane system and assume infinitely long lifetimes for the light modes $\ket{u_{i,{\bf k}}}$.  

%%%%%%%%%%%%%%%%%%%%%%%%%%%%%%%%%%%%%%%%%%%%%%%%%%%%%%%%%%%%%%%%%%%%%%%%%%%%%%
\begin{table}
\caption{\label{tab:Cherncoef}Coefficients $\alpha_{i }^{j_0}$ which, at the Dirac points, are directly related to the Chern number, as a function of $\textrm{sgn} \, M$ and $\textrm{sgn}(\sin \phi)$.}
\begin{ruledtabular}
\begin{tabular}{ccc}
& $\textrm{sgn} \, M=\textrm{sgn}\left(\sin \phi \right) $ & $\textrm{sgn} \, M=-\textrm{sgn}\left(\sin \phi \right)$ \\
\colrule
$\dfrac{\textrm{sgn}\left(\sin \phi \right)}{(-1)^i}=1$ & $ \alpha_{i}^2(\boldsymbol{K}) = - C_i$ 
&  $\alpha_{i}^1(\boldsymbol{K}') = (-1)^{i+1}C_i$\\
\noalign{\smallskip}
$\dfrac{\textrm{sgn}\left(\sin \phi \right)}{(-1)^{i+1}}=1$ & $\alpha_{i}^1(\boldsymbol{K}) = (-1)^{i} C_i $ 
&  $\alpha_{i}^2(\boldsymbol{K}') = C_i$\\
\end{tabular}
\end{ruledtabular}
\end{table}
%%%%%%%%%%%%%%%%%%%%%%%%%%%%%%%%%%%%%%%%%%%%%%%%%%%%%%%%%%%%%%%%%%%%%%%%%%%%%%

%%%%%%%%%%%%%%%%%%%%%%%%%%%%%%%%%%%%%%%%%%%%%%%%%%%%%%%%%%%%%%%%%%%%%%%%%%%%%%
\begin{table}
\caption{\label{tab:Cherncoef1}Choice of the probe's input frequency $\hbar \omega_{q_0} =E_{i,{\bf k}}$, given by the indices $i$ and ${\bf k}$, as a function of $\textrm{sgn} \, M$ and $\textrm{sgn}\left(\sin \phi \right)$, such that the transition rate depends on the Chern number: $\Gamma = J\left[E_{i,{\bf k}}\right] |C_i |/\hbar$.}
\begin{ruledtabular}
\begin{tabular}{ccc}
& $\textrm{sgn} \, M=\textrm{sgn}\left(\sin \phi \right) $ & $\textrm{sgn} \, M=-\textrm{sgn}\left(\sin \phi \right)$ \\
\colrule
$\textrm{sgn}\left(\sin \phi \right)=1$ & $i = j_0$, ${\bf k} = {\bf K} $ 
&  $i = \overline{j_0}$, ${\bf k} = {\bf K}' $\\
\noalign{\smallskip}
$\textrm{sgn}\left(\sin \phi \right)=-1$ & $i = \overline{j_0}$, ${\bf k} = {\bf K} $ 
&  $i = j_0$, ${\bf k} = {\bf K}' $\\
\end{tabular}
\end{ruledtabular}
\end{table}
%%%%%%%%%%%%%%%%%%%%%%%%%%%%%%%%%%%%%%%%%%%%%%%%%%%%%%%%%%%%%%%%%%%%%%%%%%%%%%

To acquire some intuition, let us show that the transition rate $\Gamma$ from a state $\ket{\psi(t)}$ which, at initial time $t_i$, is a probe's mode with frequency
$\omega_{q_0}$, \textit{i.e.} $\ket{\psi(t_i)} = \ket{b_{q_0}}$, to the eigenstates $\ket{u_{i,{\bf k}}}$ of the Haldane Hamiltonian's bears information about the topological character of the system. At sufficiently long times $t$, 
$\Gamma \left[\hbar \omega_{q_0} \right] = \dfrac{2 \pi}{\hbar} \sum_{i,{\bf k}} |\bra{b_{{q_0}}} H_{\textrm{cpl}} \ket{u_{i,{\bf k}}}|^2 \delta\left( \hbar \omega_{q_0} - E_{i,{\bf k}}\right)$. $\bra{b_{{q_0}}} H_{\textrm{cpl}} \ket{u_{i,{\bf k}}}$ involves the components of the Bloch state in the basis $\left( a_{1,{\bf k}}^{\dagger} , a_{2,{\bf k}}^{\dagger} \right)$ and a factor $\textrm{e}^{i {\bf k} \cdot {\bf R}_0}$ (transformation to the real space representation), such that we obtain 
$\bra{b_{{q_0}}} H_{\textrm{cpl}} \ket{u_{i,{\bf k}}} = g_{q_0} \alpha_{i }^{j_0}({\bf k})  \textrm{e}^{i {\bf k} \cdot {\bf R}_0}$.  
As one can see from Table~\ref{tab:Cherncoef}, which is constructed using Eq.~\eqref{eq:chnumber} and the related analysis of the coefficients $\alpha_{i }^{j_0}({\bf k})$, depending on $\textrm{sgn} \, M$ and $\textrm{sgn}\left(\sin \phi \right)$, it is possible to express the Chern number as a function of the coefficients $\alpha_{i }^{j_0}({\bf k})$. If $\textrm{sgn} \, M=\textrm{sgn}\left(\sin \phi \right) $ ($\textrm{sgn} \, M=-\textrm{sgn}\left(\sin \phi \right) $) we notice that the Chern number is directly related to the coefficients $\alpha_{i }^{j_0}({\bf k})$ evaluated at $\boldsymbol{K}$ ($\boldsymbol{K}'$) and $E_{i,\boldsymbol{K}}$ ($E_{i,\boldsymbol{K}'}$), $ i \in \{1,2\}$, is non-degenerate.
Therefore, choosing $\hbar \omega_{q_0} =E_{i,{\bf k}}$ with $i$ and ${\bf k}$ according to Table~\ref{tab:Cherncoef1}, we find a simple relation between $\Gamma$ and the topological invariant: $\Gamma\left(E_{i,{\bf k}}\right) = J\left(E_{i,{\bf k}}\right) |C_i |^2 =J\left(E_{i,{\bf k}}\right)|C_i |$, where the spectral function $J$ is: $J(\omega) = (2 \pi/\hbar) \sum_q g_q^2 \left[ \delta(\omega - \omega_q) -  \delta(\omega + \omega_q) \right]$. In other words, 
\begin{equation} \label{eq:gamma}
    \Gamma \left(E_{i,{\bf k}}\right) = J\left(E_{i,{\bf k}}\right) |C_i |,
\end{equation}
where the indices $i$ and ${\bf k}$ are functions of $\textrm{sgn} \, M$ and $\textrm{sgn}\left(\sin \phi \right)$ as indicated in Table~\ref{tab:Cherncoef1}. The relation appearing in Eq.~\eqref{eq:gamma} has been established from Table~\ref{tab:Cherncoef} and Table~\ref{tab:Cherncoef1}. Therefore, it relies on a fundamental property characterizing a Chern insulator: the impossibility of defining smooth Bloch states over the BZ, which translates here into the vanishing of the Bloch eigenvectors' components $\alpha_{i }^{1}$ and $\alpha_{i }^{2}$ at the opposite Dirac points.

Motivated by this, we now investigate the relation between an input voltage $ \langle V_{{\bf R}_0}^{\textrm{in}}(\omega) \rangle$ and the resulting output voltage $\langle V_{{\bf R}_0}^{\textrm{out}}(\omega) \rangle$, both at frequency $\omega$ in the probe at ${\bf R}_0$. 
For $\omega$ resolved around one Dirac point, this relation between $\langle V_{{\bf R}_0}^{\textrm{in}}(\omega)\rangle$ and $\langle V_{{\bf R}_0}^{\textrm{out}}(\omega)\rangle$ enables to rebuild the Haldane topological phase diagram.
More details on the derivation of Eq. (\ref{eq:zfregtbrij}) are given in the Supplemental Material \cite{suppl}. We indeed find
\begin{equation} \label{eq:hzfjzd}
\langle V_{{\bf R}_0}^{\textrm{out}}(\omega) \rangle= R(\omega) \langle V_{{\bf R}_0}^{\textrm{in}}(\omega)\rangle ,
\end{equation}
with $R(\omega) = 1 +i  J(\omega)\chi_{{\bf R}_0,{\bf R}_0} $,
and
\begin{widetext}
\begin{equation} \label{eq:xi}
\chi_{{\bf R}_0,{\bf R}_0} = \dfrac{1}{ N} \sum_{i=1}^2 \sum_{{\bf k} } \gamma_{j_0,{\bf k}}^{i} \left[ \dfrac{1}{- \hbar \omega - E_{i,{\bf k}} + i 0^+} - \dfrac{1}{-\hbar \omega + E_{i,{\bf k}} + i 0^+} \right],
\end{equation}
\end{widetext}
where $N$ is the number of lattice sites
and 
\begin{equation} \label{eq:zfregtbrij}
\gamma_{j_0,{\bf k}}^{i} = \dfrac{1}{2}+\dfrac{(-1)^{j_0+i+1} h_2({\bf k})}{2 \epsilon({\bf k})} \in \, \mathbb{R}.
\end{equation}
The key point within our present approach is to observe that $\gamma_{j_0,{\bf k}}^{i}$ evaluated at the Dirac points, where $h_1=0$, depends only on the sign of the function $h_2$: we have $2 \gamma_{j_0,{\bf k}}^{i} = 1 - (-1)^{j_0+i} \textrm{sgn} \, h_2({\bf k}) $ for ${\bf k} = \{{\bf K}, {\bf K}'\}$. The response function is then directly related to the topological invariant via Eq.~\eqref{eq:chnumber} and Table~\ref{tab:sgnh2}. As we show in Table~\ref{tab:alphabeta}, depending on the sign of $\sin \phi $ and on the sign of the Semenoff mass, the $i^{\textrm{th}}$ band topological invariant is given by the coefficient $\gamma_{j_0,{\bf k}}^{i}$, evaluated at $j_0 =i$ or $j_0 =\overline{i}$ and at ${\bf k} = {\bf K}$ or ${\bf k} = {\bf K}'$, with $\overline{i} = 2(1)$ if $i=1(2)$. Again, this outcome arises from a fundamental characteristic associated to the topological phase: the vanishing of the Bloch eigenvectors' components $\alpha_{i }^{1}$ and $\alpha_{i }^{2}$ at the opposite Dirac points. 

%%%%%%%%%%%%%%%%%%%%%%%%%%%%%%%%%%%%%%%%%%%%%%%%%%%%%%%%%%%%%%%%%%%%%%
\begin{table}[b]
\caption{\label{tab:alphabeta} Value of $\gamma_{j_0,{\bf k}}^{i}$, evaluated at $j_0 =i$ or $j_0 =\overline{i}$ and at ${\bf k} = {\bf K}$ or ${\bf k} = {\bf K}'$, as a function of $\textrm{sgn}\left(\sin \phi \right)$ and $\textrm{sgn} \, M$. We remind that $\overline{i} = 2(1)$ if $i=1(2)$.}
\begin{ruledtabular}
\begin{tabular}{ccc}
& $\textrm{sgn} \, M=\textrm{sgn}\left(\sin \phi \right) $ & $\textrm{sgn} \, M=-\textrm{sgn}\left(\sin \phi \right)$ \\
\colrule \noalign{\smallskip}
$\textrm{sgn}\left(\sin \phi \right)=1$ & $ \gamma_{i,\boldsymbol{K}}^i = (-1)^{\overline{i}} C_i$ 
&  $\gamma_{\overline{i},\boldsymbol{K}'}^{i} = (-1)^{\overline{i}} C_i$\\
\noalign{\smallskip}
$\textrm{sgn}\left(\sin \phi \right)=-1$ & $\gamma_{\overline{i},\boldsymbol{K}}^{i} = (-1)^{i} C_i$ 
&  $\gamma_{i,\boldsymbol{K}'}^i = (-1)^{i} C_i$\\
\end{tabular}
\end{ruledtabular}
\end{table}
%%%%%%%%%%%%%%%%%%%%%%%%%%%%%%%%%%%%%%%%%%%%%%%%%%%%%%%%%%%%%%%%%%%%

We can now understand how measuring the reflected light signal in the probe reveals the topological phase transition of the two-dimensional lattice model.
We write $S^{\textrm{in}}(\omega)=\left| \langle V_{{\bf R}_0}^{\textrm{in}}(\omega)\rangle\right|^2$ and $S^{\textrm{out}}(\omega)=\left| \langle V_{{\bf R}_0}^{\textrm{out}}(\omega)\rangle\right|^2$ the energy spectral density respectively associated to the input and output voltages.
To leading order in the coupling amplitudes, we have $S^{\textrm{out}}(\omega) = | R(\omega) |^2 S^{\textrm{in}}(\omega)$ and for $\omega>0$,
\begin{equation}\label{eq:realRrrreaf}
  |R(\pm \omega) |^2 =  1 \mp \dfrac{2\pi J(\pm \omega)}{N}   \sum_{i=1}^2 \sum_{{\bf k} }  \gamma_{j_0,{\bf k}}^{i} \delta( \hbar \omega - E_{i,{\bf k}}).
\end{equation}
For $\textrm{sgn} \, M=\textrm{sgn}\left(\sin \phi \right) $, the energies $E_{i,\boldsymbol{K}}, \, i \in \{1,2\}$ are non-degenerate, therefore, choosing $\hbar \omega = E_{i,\boldsymbol{K}}$ selects only the ${\bf k} = {\bf K}$ point in the integral appearing in the Eq.~\eqref{eq:realRrrreaf}.  
Moreover, as indicated in Table~\ref{tab:alphabeta},
$\gamma_{j_0,{\bf K}}^{i}$ is related to the topological invariant if we choose a probe at $j_0=i$ ($\overline{i}$) for $\textrm{sgn}\left(\sin \phi \right)=1$ ($\textrm{sgn}\left(\sin \phi \right)=-1$). Therefore, for a well-chosen frequency $\omega$, $|R(\omega) |^2 $ clearly depends on the topological invariant. This is also true for $\textrm{sgn} \, M=-\textrm{sgn}\left(\sin \phi \right) $, if, in the previous analysis, we replace ${\bf K}$ by ${\bf K}'$ and $j_0$ by $\overline{j_0}$. 

\paragraph*{ Finite lifetimes for the light modes.---} Eventually, we address the more realistic scenario in which we incorporate finite lifetimes for both the modes in the probe $\ket{b_{q}}$ and the Chern insulator's modes $\ket{u_{i,{\bf k}}}$. For simplicity, we consider the same bandwidth amplitude $\Delta_{\textrm{CI}}$ ($\Delta_{\textrm{P}}$) for all the modes $\ket{u_{i,{\bf k}}}$ ($\ket{b_{q}}$). 
We assume the following ordering of the energies $\textrm{max}_q(g_q) \ll \{\Delta_{\textrm{CI}},\Delta_{\textrm{P}}\} \ll \{t_1, \hbar\textrm{min}_{q,q'}(|\omega_{q}-\omega_{q'}|)\}$. 
We replace the Dirac Delta functions appearing in Eq.~\eqref{eq:realRrrreaf} by normalized Gaussian spectral distributions denoted $G\left(\omega;\overline{\omega},\Delta\right)$ with mean value $\overline{\omega}$ and standard deviation $\Delta$: 
$\delta( \hbar \omega - E_{i,{\bf k}})$ is replaced by $G\left(\omega;E_{i,{\bf k}}/\hbar, \Delta_{\textrm{CI}}\right)$ and 
$J(\omega)$ is replaced by $\tilde{J}[\omega] = 2 \pi \sum_q \left(g_q/\hbar\right)^2 \left[ G\left(\omega,\omega_q,\Delta_{\textrm{P}}\right)-  G\left(\omega,-\omega_q,\Delta_{\textrm{P}}\right) \right]$.
We also consider an input energy spectral density with Gaussian distribution: $S^{\textrm{in}}(\omega) = G\left(\omega;\omega_{q_0},\Delta_{\textrm{P}}\right) $. 
For a well chosen $\omega_{q_0}$ ($E_{i,\boldsymbol{K}}$ or $E_{i,\boldsymbol{K}}$), $|R(\omega) |^2 $ still depends on the topological invariant because $\gamma_{j_0,{\bf k}}^{i}$ is directly related to the Chern number.
It leads to a decrease of the output peak's weight $\int d \omega S^{\textrm{out}}(\omega)$ compared to the normalized weight of the input peak. This decrease is given by
\begin{equation} \label{eq:output}
  1-\int d \omega S^{\textrm{out}}(\omega) =  \dfrac{2\pi }{N}   \sum_{i=1}^2 \sum_{{\bf k} }  \gamma_{j_0,{\bf k}}^{i} I_{i,{\bf k}},
\end{equation}
with $I_{i,{\bf k}}=\int d \omega \tilde{J}(\omega) G\left(\omega;E_{i,{\bf k}}/\hbar, \Delta_{\textrm{CI}}\right) G\left(\omega;\omega_{q_0},\Delta_{\textrm{P}}\right) $, which is $\left( g_{q_0}/\hbar\right)^2 /\left( \sqrt{2 \pi }\Delta_{\textrm{CI}} \Delta_{\textrm{P}}^2\right) $ times the overlap area $\int d\omega \, \textrm{exp}{- \dfrac{\left(\omega- \omega_{q_0} \right)^2}{ \Delta_{\textrm{P}}^2 } } \, \textrm{exp} {- \dfrac{\left(\omega-E_{i,{\bf k}}/\hbar\right)^2}{2 \Delta_{\textrm{CI}}^2 }}$ . 

\paragraph*{Disorder.---}
We expect the general structure of the wave function over space, which is related to the bulk invariant, to be robust against weak disorder. This central feature gives robustness to the probe proposed in this letter. To illustrate this point, we consider a finite size system with local disorder on the parameters. We do not have translational symmetry but the expression of $1-\int d \omega S^{\textrm{out}}(\omega)$ in Eq.~\eqref{eq:output} is easily adapted: the sum over $i$ and $\bf k$ is replaced by a sum over the lattice sites and a numerical diagonalization gives the energies and the decomposition of the eigenvectors over the lattice sites from which we get the analogues of the coefficient $\gamma_{j_0,{\bf k}}^{i}$ and the integral $I_{i,{\bf k}}$.
 
For one disorder configuration, we choose to sample each of the values of $t_1$, $t_2$, $M$ and $\phi$ over the lattice from a Gaussian distribution law with respectively mean value $\overline{t}_1$, $\overline{t}_2$, $\overline{M}$ and $\overline{\phi}$ and standard deviation being five percent of the associated mean value. We choose the experimentally relevant energy scales $\Omega_0=10$ GHz, $\overline{t}_1/\hbar=100$ MHz, $\overline{t}_2/\hbar=15 $ MHz, $\Delta_{\textrm{CI}} = \Delta_{\textrm{P}} = 10$MHz and $g_{q_0}/\hbar=1$MHz. These scales correspond to a relatively low quality factor $Q=\Omega_0/\Delta_{\textrm{CI}} =10^3$ (here the same for both the cQED Chern insulator and the probe) and a low coupling amplitude and should be reachable in a cQED experiment. In Fig.~\ref{fig1}(d), we show a numerical evaluation of $1-\int d \omega S^{\textrm{out}}(\omega)$ as a function of $\overline{M}/\overline{t}_2$ and $\overline{\phi}$. 
For this figure, we consider $ \overline{M}>0$, the probe is coupled to a sublattice A site ($j_0 =1$) and $\sin \overline{\phi} >0$. Note that if the latter inequality is arbitrarily imposed then our measure can not access the sign of the Chern number, but it still discriminates between a topological and a trivial phase. Moreover, we chose $\hbar \omega_{q_0} = \tilde{E}_{1,\boldsymbol{K}}$ and we expect $1-\int d \omega S^{\textrm{out}}(\omega)$ depends on $ C_1 $ (because $\gamma_{1,{\bf K}}^1 = C_1 $). $\tilde{E}_{1,\boldsymbol{K}}$ is the highest energy of the lowest band and it can be determined through the energy density of states before the measure of the Chern number. From Eqs.~\eqref{eq:zfregtbrij} and \eqref{eq:realRrrreaf}, we observe that the energy density of states is obtained by summing the local responses to an input measured in two distinct probes on sublattices A and B. In our protocol, $\textrm{sgn} \, M$ also needs to be determined before the measure of the Chern number.

%\paragraph*{Magnon system and material suggestions.---} The results described here are directly relevant for other Chern insulator systems. For instance, our probe can be used for a topological magnon insulator, as the one proposed in Ref.~\cite{Owerre16}, which is described by a bosonic Haldane model. Indeed, consider a magnetic tip described by a polarization state denoted ${\bf s}=(s^x,s^y,s^z)$, and assume the tip thin enough to couple to only one magnetic site ${\bf S}_i$ of the system. The Hamiltonian coupling the tip to the magnon system is $H_{\textrm{cpl}} = J \, {\bf S}_i \cdot {\bf s}_i$, with $J$ the coupling amplitude. Suppose now the tip is polarized along $x$ only. Then $H_{\textrm{cpl}} = J \, {\bf S}_i \cdot {\bf s}_i = (J/4) (S_i^+ + S_i^-) (s^+ + s^-)$, where $S_i^+$, $S_i^-$, $s^+$ and $s^-$ are bosonic creation/annihilation operators. $H_{\textrm{cpl}}$ is completely analogue to the coupling Hamiltonian we considered for the cQED system. Therefore, measuring the response to a magnetic excitation provides a way to access the topological number. Such a measurement should be applicable in topological magnon quantum materials, such as CrI$_3$ \cite{Chen18} or maybe $\beta$-Cu$_2$V$_2$O$_7$ \cite{Tsirlin10}. Even though the magnon can condense at the lowest energy mode, thermal excitation shall generate magnon modes at all energies in the system, namely the ones required for the proposed probe, with energy around the Dirac point.

\paragraph*{ Remarks.---} Two observations are in order.

(i) The probe is able to measure the topological number based on a real space local coupling to the system and with a resolution in reciprocal space thanks to the energy conservation, similarly to circularly polarized light \cite{HurK22}. 

(ii) If the input is triggered at a site identified by $\left( {\bf r}, j_0\right)$ and the output is measured at $\left( {\bf r}', j_0'\right)$, the expression in the summation of Eq.~\eqref{eq:xi} becomes
\begin{equation} \label{eq:rq}
    \dfrac{\left[\beta_{j_0}^i({\bf k}) \alpha_{i}^{j_0'}({\bf k})\right]^* \textrm{e}^{i {\bf k}\left({\bf r}-{\bf r}'\right)}}{-\hbar \omega - E_{i,{\bf k}} + i 0^+} - \dfrac{\beta_{j_0}^i({\bf k}) \alpha_{i}^{j_0'}({\bf k}) \textrm{e}^{-i{\bf k} \left({\bf r}-{\bf r}'\right)}}{-\hbar \omega + E_{i,{\bf k}} + i 0^+} ,
\end{equation}
with $\beta_{j_0 }^{i}({\bf k}) \alpha_{i }^{j_0}({\bf k})=\gamma_{j_0,{\bf k}}^{i}$ and for $j_0' =\overline{j_0} \neq j_0$, $\beta_{j_0}^i({\bf k}) \alpha_{i}^{\overline{j_0}}({\bf k}) \propto h_1({\bf k}) / 2 \epsilon({\bf k})$.
At the Dirac points, $h_1$ is vanishing, therefore, in the case $j_0' =\overline{j_0}$, the simple protocol we sketched above does not help to rebuild the topological phase diagram. This outcome can be anticipated based on the fact that, at one given Dirac point, one of both Bloch eigenvectors' components vanish, in the topological regime. In the scenario $j_0' =j_0$,  because the coefficients in the numerator of Eq.~\eqref{eq:rq} are complex-valued, the Chern number dependency of $1-\int d \omega S^{\textrm{out}}(\omega)$ is mitigated. Indeed, the latter contains principal values of integrals over frequency involving $\left[1/\left(-\hbar \omega \pm E_{i,{\bf k}}\right)\right]$ terms.

\paragraph*{ Conclusion.---} We have introduced a local microwave-light probe with capacitive coupling to a cQED array described by a Haldane bosonic system, in the regime of small coupling amplitudes. We have explained how this probe is relevant for the detection of the topological character of Chern insulators. Using Fermi golden rule, we established a connection between the Chern number and the transition rate from a probe's eigenstate (with frequency corresponding to one of the Dirac points energy) to the eigenstates of the Haldane Hamiltonian. Secondly, we developed the input-output theory for the probe, enabling us to compute the reflection coefficient which relates an input voltage and an output voltage. We showed that for an input with frequency resolved at one of the Dirac points, this reflection coefficient is directly related to the system's topological invariant. The fundamental working principle of this probe makes it inherently relevant for Chern insulators with non-degenerate Dirac points in the energy spectrum (for the Haldane system we consider this imposes $t_2 \lesssim 0.15t_1$). As a future prospect, it appears intriguing to adapt this probe to other systems that may exhibit different particle statistics, such as cold atoms or various material platforms.

\begin{acknowledgments}
This work was supported by the french ANR BOCA grant. JL acknowledges support from the National Research Fund Luxembourg under Grant No. INTER/QUANTERA21/16447820/MAGMA.
\end{acknowledgments}

\bibliography{ref}

\clearpage
\input{sup.tex}

\end{document}

%% file: sup.tex
\renewcommand{\thetable}{S\Roman{table}}
\setcounter{table}{0}  

\renewcommand{\theequation}{S\arabic{equation}}
\setcounter{equation}{0}  

\setcounter{secnumdepth}{2}

\renewcommand{\thesection}{S\arabic{section}}
\renewcommand{\thesubsection}{\thesection.\arabic{subsection}}
\renewcommand{\thesubsubsection}{\thesubsection.\arabic{subsubsection}}

\phantomsection
\begin{large}
\begin{center}
    \textbf{Supplemental Material} 
\end{center}
\end{large}

In this Supplemental Material, in Sec.~\ref{app:Haldane}, we derive the formula $C_{j} =(-1)^{j+1} \, \textrm{sgn}( \sin \phi )$ for the topological phase using standard geometrical definitions.
In Sec.~\ref{app:inputoutput}, we give additional information on the local probe and on the derivation of Eq. (6) in the Letter.

\section{Chern number and Berry gauge fields} \label{app:Haldane}

In this section we show a detailed analytical calculation of the Chern number, based on an approach introduced by Kohmoto \cite{Kohmoto85}, for the Haldane Hamiltonian. 
This computation relies on the definition of two distinct gauge choices $G_I$ and $G_{II}$ for the Bloch eigenvectors $
\ket{u_{i,{\bf k}}} = \left[\alpha_{i}^{1}({\bf k}) a_{1,{\bf k}}^{\dagger} + \alpha_{i}^{2}({\bf k}) a_{2,{\bf k}}^{\dagger} \right] \ket{0} \,,
$ that we respectively write $\ket{u_{i,{\bf k},I}}$ and $\ket{u_{i,{\bf k},II}}$.

(i) \textit{Gauge choice $G_I$}: the coefficient $\alpha_{i }^{1}({\bf k})$ is real. 
More precisely, we choose 
\begin{equation} \label{eig0}
\alpha_{i }^{1}({\bf k})=  \rho_i({\bf k}) \,,
\end{equation}
with 
\begin{equation}
\rho_i({\bf k}) = \dfrac{h_2({\bf k}) + (-1)^i \epsilon({\bf k}) }{2\left[\epsilon({\bf k})^2 +(-1)^i h_2({\bf k}) \epsilon({\bf k}) \right]^{1/2}},
\end{equation}
and we have
\begin{equation}
\alpha_{i }^{2}({\bf k})  =  \lambda_i({\bf k}) \textrm{e}^{-i \varphi({\bf k})} \, ,
\end{equation}
 with 
 \begin{equation}
 \lambda_i({\bf k}) = \dfrac{|h_1({\bf k}) |}{2\left[\epsilon({\bf k})^2 +(-1)^i h_2({\bf k}) \epsilon({\bf k}) \right]^{1/2}} ,
 \end{equation}
 and
\begin{equation} 
\textrm{e}^{-i \varphi({\bf k})}=\dfrac{ \textrm{e}^{i {\bf k} \cdot {\bf a}_3} h_1({\bf k})^* }{ |h_1({\bf k})| }.
\end{equation}

(ii) \textit{Gauge choice $G_{II}$}: the coefficient $ \alpha_{i }^{2}({\bf k})$ is real. 
We choose $ \ket{u_{j,{\bf k},II}}=\textrm{e}^{i \varphi({\bf k})}\ket{u_{j,{\bf k},I}}$, \textit{i.e.}
\begin{equation}
\alpha_{i }^{2}({\bf k}) =\lambda_i({\bf k}),
\end{equation}
and we have
\begin{equation}
\alpha_{i }^{1}({\bf k}) = \rho_i({\bf k}) \textrm{e}^{i \varphi({\bf k})}.
\end{equation}

%%%%%%%%%%%%%%%%%%%%%%%%%%%%%%%%%%%%%%%%%%%%%%%%%%%%%%%%%%%%%%%%%%%%%%
\begin{table}[b]
\caption{\label{tab:rho1}Table giving the values of of $\rho_i({\bf k})$ and $ \lambda_i({\bf k})$ at the Dirac points in the case $|M|< 3 \sqrt{3} t_2 |\sin \phi|$ , as a function of the sign of the Semenoff mass $M$ and of the index of the energy band $i$.}
\begin{tabular}{ccccc}
\hline
\hline
 &  $\textrm{sgn}( \sin \phi ) =(-1)^i \, $ & $\textrm{sgn}( \sin \phi ) =(-1)^{i+1} \, $  \\
\hline
$\rho_i({\bf k} \rightarrow {\bf K}) $ & $0$ 
&  $(-1)^i$ \\
$ \rho_i({\bf k} \rightarrow {\bf K}')$ &  $(-1)^i$  &  $0$\\
$  \lambda_i({\bf k} \rightarrow {\bf K}) $  &  $1$  &  $0$\\
$ \lambda_i({\bf k} \rightarrow {\bf K}')$ & $0$ &  $1$ \\
\hline
\hline
\end{tabular}
\end{table}
%%%%%%%%%%%%%%%%%%%%%%%%%%%%%%%%%%%%%%%%%%%%%%%%%%%%%%%%%%%%%%%%%%%%
%%%%%%%%%%%%%%%%%%%%%%%%%%%%%%%%%%%%%%%%%%%%%%%%%%%%%%%%%%%%%%%%%%%%%%
\begin{table}[b]
\caption{\label{tab:rho3}Table giving the values of of $\rho_i({\bf k})$ and $ \lambda_i({\bf k})$ at the Dirac points in the case $|M| > 3 \sqrt{3} t_2 |\sin \phi|$, as a function of the sign of the Semenoff mass $M$ and of the index of the energy band $i$.}
\begin{tabular}{ccccc}
\hline
\hline
 &  $\textrm{sgn}\, M  =(-1)^i \, $ & $\textrm{sgn}\, M  =(-1)^{i+1} \, $  \\
\hline
$\rho_i({\bf k} \rightarrow {\bf K}) $ & $(-1)^i$ 
&  $0$ \\
$ \rho_i({\bf k} \rightarrow {\bf K}')$ &  $(-1)^i$  &  $0$\\
$  \lambda_i({\bf k} \rightarrow {\bf K}) $  &  $0$  &  $1$\\
$ \lambda_i({\bf k} \rightarrow {\bf K}')$ & $0$ &  $1$ \\
\hline
\hline
\end{tabular}
\end{table}
%%%%%%%%%%%%%%%%%%%%%%%%%%%%%%%%%%%%%%%%%%%%%%%%%%%%%%%%%%%%%%%%%%%%

The phase of the wavefunction is chosen by requiring that either the coefficient $\alpha_{i }^{1}({\bf k})$ or the coefficient $ \alpha_{i }^{2}({\bf k}) $ is real (and non-zero). 
$\rho_i({\bf k})$ and $ \lambda_i({\bf k})$ may vanish only at the Dirac points and it depends on the sign of $h_2$. For $|M| <3 \sqrt{3} t_2 |\sin \phi|$ ($|M| >3 \sqrt{3} t_2 |\sin \phi|$), the values of $\rho_i({\bf k})$ and $ \lambda_i({\bf k})$ at the Dirac points are given in Table~\ref{tab:rho1} (Table~\ref{tab:rho3}).
 
(i) In the case $|M|< 3 \sqrt{3} t_2 |\sin \phi|$, we notice that (see Table~\ref{tab:rho1}), for all the values of $\phi$, none of both gauge choices $G_{I}$ and $G_{II}$ is well-defined in the entire BZ. Indeed, $\rho_i({\bf k})$ and $\lambda_i({\bf k})$ each vanish at one of the Dirac points. 
We then define two non-overlapping domains $\mathcal{D}_I$ and $\mathcal{D}_{II}$ in the BZ, each containing a different Dirac point, and we use a different gauge choice for the Bloch states in each domain \cite{Kohmoto85}. To be more specific, for $(-1)^i \, \textrm{sgn}( \sin \phi ) = +1$, we apply $G_I$ for the points contained in $\mathcal{D}_{II}$ and $G_{II}$ for the points contained in $\mathcal{D}_{I}$ while for $(-1)^i \, \textrm{sgn}( \sin \phi ) = -1$, we apply $G_{I}$ for the points contained in $\mathcal{D}_{I}$ and $G_{II}$ for the points contained in $\mathcal{D}_{II}$.
Then the phase of $\ket{u_{i,{\bf k},I}}$ (the eigenstate in $\mathcal{D}_{I}$) and $\ket{u_{i,{\bf k},II}}$ (the eigenstate in $\mathcal{D}_{II}$) and the Berry gauge fields ${\bf A}_{i,{\bf k},I} = \bra{u_{i,{\bf k},I}}{\bf \nabla}_{{\bf k}} \ket{u_{i,{\bf k},I}}$ and ${\bf A}_{i,{\bf k},II} = \bra{u_{i,{\bf k},II}}{\bf \nabla}_{{\bf k}} \ket{u_{i,{\bf k},II}}$ are uniquely and smoothly defined respectively on $\mathcal{D}_I$ and $\mathcal{D}_{II}$. 
The Chern number associated to the $j^{\textrm{th}}$ band (remind that $j=1$ or $j =2$) reads
\begin{align} 
\begin{split}
\label{chern_nunununu}
    C_j = \dfrac{1}{2 i \pi} \bigg[ &\int_{\mathcal{D}_{\textrm{I}}} d^2{\bf k} \cdot \left( {\bf \nabla}_{{\bf k}} \times {\bf A}_{j,{\bf k},\textrm{I}} \right) \\ &+\int_{\mathcal{D}_{\textrm{II}}} d^2{\bf k} \cdot \left( {\bf \nabla}_{{\bf k}} \times {\bf A}_{j,{\bf k},\textrm{II}} \right)\bigg],
\end{split}
\end{align}
where $d^2{\bf k}$ is an oriented infinitesimal surface element and ${\bf \nabla}_{{\bf k}} \times {\bf A}_{j,{\bf k},\textrm{I}}$ and ${\bf \nabla}_{{\bf k}} \times {\bf A}_{j,{\bf k},\textrm{II}}$ are the Berry curvatures respectively associated to ${\bf A}_{j,{\bf k},\textrm{I}}$ and ${\bf A}_{j,{\bf k},\textrm{II}}$. Let us define a closed path P along the boundary between $\mathcal{D}_I$ and $\mathcal{D}_{II}$, surrounding once the Dirac points. Using Stokes' theorem leads to
\begin{equation} 
	C_j =  \dfrac{1}{2 i \pi} \left( \oint_{\textrm{P}} d{\bf k}  \cdot {\bf A}_{j,{\bf k},\textrm{I}} - \oint_{\textrm{P}} d{\bf k} \cdot {\bf A}_{j,{\bf k},\textrm{II}}\right).
\end{equation}
$\oint_{\textrm{P}} d{\bf k}$ is a line integral along P, where we have $\ket{u_{j,{\bf k},I}} =\textrm{e}^{i (-1)^j \, \textrm{sgn}( \sin \phi ) \varphi({\bf k})}\ket{u_{j,{\bf k},II}}$. 
We choose $\varphi({\bf k})$ so that it is smooth along the whole P path and we obtain
\begin{equation} 
	C_{j} = \dfrac{(-1)^{j} \,\textrm{sgn}( \sin \phi )}{2 \pi} \oint_{\textrm{P}} d{\bf k} \cdot {\bf \nabla}_{{\bf k}} \varphi ({\bf k}).
\end{equation}

$C_i$ is found by studying how $\varphi({\bf k})$ evolves when moving along P. Generally speaking, when the P path surround a $\varphi ({\bf k})$'s divergence (here at the ${\bf K}$ or ${\bf K}'$ point), the accumulated phase increases or decreases by $\pm 2 \pi z, \, z \in \mathbb{Z}$, which gives a quantized value of $C$, as expected. Here, we find that $\varphi ({\bf k})$ changes by $-2 \pi$ when moving along the entire closed path P. This gives 
\begin{equation}
C_{j} =(-1)^{j+1} \, \textrm{sgn}( \sin \phi ). 
\end{equation}
One can build intuition from small displacement $\delta {\bf k}$ around $\bf K$, in which case we have $h_1({\bf K} + \delta {\bf k}) = 3 t_1(- \delta k_x + i \delta k_y)/2$ where $\delta k_x = \delta {\bf k} \cdot {\bf e}_x$, $\delta k_y = \delta {\bf k} \cdot {\bf e}_y$, ${\bf e}_x=-\left({\bf b}_2+ {\bf b}_3 \right)/\sqrt{3}$ and ${\bf e}_y=\left({\bf b}_2- {\bf b}_3 \right)/3$. In this case, we easily see that $\varphi$, which is defined by $\textrm{e}^{-i \varphi({\bf k})}=\textrm{e}^{i {\bf k} \cdot {\bf a}_3} h_1({\bf k})^* / |h_1({\bf k})| $, varies by $-2 \pi$, for one entire closed path P around $\bf K$ (oriented anticlockwise).
As mentioned in the letter, this form is equivalent to the one in Eq. (2) which is local at the Dirac points.

(iii) In the case $|M| > 3 \sqrt{3} t_2 |\sin \phi|$, we gathered the values of $\rho_i({\bf k})$ and $ \lambda_i({\bf k})$ at the Dirac points in Table~\ref{tab:rho3}. At both Dirac points, either $\rho_i({\bf k})$ or $\lambda_i({\bf k})$ are non-vanishing, therefore it is possible to find a unique and smooth phase for the ket $\ket{u_{i,{\bf k}}}$ everywhere in the BZ which leads to a unique and smooth Berry gauge field ${\bf A}_{{\bf k}}$. Depending on the value of $(-1)^i \, \textrm{sgn}( M)$, we apply gauge choice $G_{I}$ or $G_{II}$ for all the points of the BZ, and then we can show that the associated wave function $\ket{u_{i,{\bf k},I}}$ or $\ket{u_{i,{\bf k},II}}$ (and its phase) is uniquely and smoothly defined, as is the Berry gauge field. Because the BZ is a torus, the Chern numbers $C_j$ are vanishing.

\section{Local response to capacitively coupled probes in a two-dimensional lattice bosonic system}\label{app:inputoutput}

In this Section, we consider a set of bosonic probes (typically microwave light resonators) capacitively coupled to a bosonic lattice model. We derive the relation between the output voltage operator in a probe at a certain position on the lattice and the input voltage operators associated to the ensemble of probes on the lattice. We consider a 2-dimensional lattice system with periodic boundary conditions. 

\subsection{Hamiltonian} The Hamiltonian describing the lattice model with capacitively coupled probes reads
\begin{equation}
	H =  H_{\textrm{lat}} + H_{\textrm{prb}} + H_{\textrm{cpl}}, 
\end{equation}
where $H_{\textrm{lat}}$ is the (topological) lattice Hamiltonian, $H_{\textrm{prb}}$ is the Hamiltonian associated to the probe(s) and $H_{\textrm{cpl}}$ is the Hamiltonian associated to the coupling between the lattice and the probe(s).

Let us call $N_C$ the number of sites per unit cell in the lattice we consider and $N$ the total number of unit cells. We label each site within a unit cell with different colors, and two different sites belonging to the same Bravais lattice are labeled by the same color. 
We define the Fourier transform of the annihilation operator of a (bosonic) particle at position $\bf r$ and the inverse relation 
\begin{equation}
a_{j,{\bf k}} = \sum_{{\bf r} \in R_{j}} \textrm{e}^{-i {\bf k} .{\bf r}} a_{{\bf r}} \quad \textrm{and} \quad a_{{\bf r}} = \dfrac{1}{N} \sum_{{\bf k}} \textrm{e}^{i {\bf k} .{\bf r}} a_{j({\bf r}),{\bf k}},
\end{equation}
with $R_j$ the ensemble containing all the lattice positions of the color-$j$ sites and the function $j({\bf r})$ returns the color index at the ${\bf r}$ site.
We formally write
\begin{equation}
	H_{\textrm{lat}} = \sum_{\boldsymbol{k}} \Psi_{\boldsymbol{k}}^\dagger h_{\boldsymbol{k}} \Psi_{\boldsymbol{k}},
\end{equation}
with $\Psi_{{\bf k}}^\dagger = \left(a_{1,{\bf k}}^\dagger, \dots a_{N,{\bf k}}^\dagger\right)$. We write $h_{{\bf k}}$'s associated eigenvalues $E_{i,{\bf k}}$, where $i \in \{1, \dots, N_C\}$, and we write the associated eigenvectors 
\begin{equation}
\ket{\Phi_{i,{\bf k}}} = \Phi_{i,{\bf k}}^\dagger \ket{0} = \sum_{j=1}^{N_C} \alpha_{i }^{j}({\bf k}) a_{j,{\bf k}}^\dagger \ket{0},
\end{equation}
with $\alpha_{i }^{j}({\bf k}) \in \mathbb{C}$. We have
\begin{equation}
	H_{\textrm{lat}} = \sum_{{\bf k}} \sum_{i=1}^{N_C} E_{i,{\bf k}} \Phi_{i,{\bf k}}^\dagger \Phi_{i,{\bf k}}.
\end{equation}
Each probe are resonators with a certain number of relevant  modes, each mode $q$ being characterized by the frequency $\omega_q$. Therefore we write
\begin{equation}
H_{\textrm{prb}} = \sum_{{\bf r} \in R_p} \sum_q \hbar \omega_q b_{\boldsymbol{r},q}^\dagger b_{{\bf r},q},
\end{equation}
with $b_{\boldsymbol{r},q}$ the annihilation operators for the mode $q$ of the probe at $\boldsymbol{r} \in R_p$, $R_p$ being the ensemble of the positions of the nodes coupled to a probe.
We assume a capacitive coupling between each probe and a node of the lattice. The Hamiltonian reads
\begin{equation}
H_{\textrm{cpl}} = \sum_{{\bf r} \in R_p} \left( a_{\bf r}+ a_{{\bf r}}^\dagger \right)\sum_q g_q \left( b_{{\bf r},q}+ b_{{\bf r},q}^\dagger  \right).
\end{equation}
The coupling amplitude $g_q$ is assumed not to depend on the position of the probe. 

\subsection{Input-output analysis} 

Here we use the input-output formalism as reviewed in Ref.~\cite{Clerk10}.

Let us define the input voltage in the probe at $\bf r$
\begin{equation}
V_{\bf r}^{\textrm{in}}(t) =\sum_q g_q \left[ \textrm{e}^{-i \omega_q (t-t_i)}b_{{\bf r},q}(t_i) + h.c. \right],
\end{equation}
where $t_i < t$ is an initial time in the distant past, and the output voltage in the probe at $\bf r$
\begin{equation}
V_{\bf r}^{\textrm{out}}(t) = \sum_q g_q \left[ \textrm{e}^{-i \omega_q (t-t_f)} b_{{\bf r},q}(t_f) + h.c. \right],
\end{equation}
where $t_f>t$ is a final time in the distant future. Let us call $x_{\bf r} =  a_{\bf r}+ a_{{\bf r}}^\dagger$. The Heisenberg equation of motion (EOM) for the $b_{jq}$ operator reads\begin{equation} \label{eomb}
	\dot{b}_{{\bf r},q} = \dfrac{i}{\hbar}  \left[H,b_{{\bf r},q}\right] = -i \omega_q b_{{\bf r},q} - \dfrac{i  g_q}{\hbar} x_{\bf r}.
\end{equation}
The solution of this equation of motion is
\begin{equation} \label{fi0}
	 b_{{\bf r},q} (t) = \textrm{e}^{-i \omega_q (t-t_i)}b_{{\bf r},q}(t_i) - \dfrac{i  g_q}{\hbar} \int_{t_i}^t d\tau \textrm{e}^{-i \omega_q (t-\tau)} x_{\bf r}(\tau),
\end{equation}
or equivalently
\begin{equation}
	 b_{{\bf r},q} (t) = \textrm{e}^{-i \omega_q (t-t_f)}b_{{\bf r},q}(t_f) + \dfrac{i  g_q}{\hbar} \int_{t}^{t_f} d\tau \textrm{e}^{-i \omega_q (t-\tau)} x_{\bf r}(\tau).
\end{equation}
Combining the previous equations and their complex conjugate counterparts we get 
\begin{equation} \label{nan}
V_{\bf r}^{\textrm{in}}(t) - 2 \sum_q \dfrac{g_q^2}{\hbar} \int_{t_i}^{t_f} d \tau \sin \left[ \omega_q (t- \tau)\right] x_{\bf r}(\tau) = V_{\bf r}^{\textrm{out}}(t).
\end{equation}
We Fourier transform the previous equation with respect to the time variable $t$, we define $J(t) = 2 i \sum_q \dfrac{g_q^2}{\hbar} \sin \left( \omega_q t\right) $ \cite{Schiro14} and we get
\begin{equation} \label{ano}
V_{\bf r}^{\textrm{out}}(\omega) = V_{\bf r}^{\textrm{in}}(\omega) +i J(\omega) x_{\bf r}(\omega).
\end{equation}
Notice that $J(\omega) \in \mathbb{R}$. Its explicit expression is
\begin{equation}
J(\omega) = 2 \pi \sum_q \dfrac{g_q^2}{\hbar} \left[ \delta(\omega - \omega_q) - \delta(\omega + \omega_q) \right],
\end{equation}
where $\delta$ is the Dirac delta function.

Now we express $x_{\bf r}(\omega)$ as a function of $V_{\bf r}^{\textrm{in}}(\omega)$. The Heisenberg EOM for the $\Phi_{i,{\bf k}}$ modes reads
\begin{align} \label{eom}
\begin{split}
   \hbar \dot{\Phi}_{i,{\bf k}} = &-i E_{i,{\bf k}} \Phi_{i,{\bf k}} \\ &- i \sum_{{\bf r} \in R_p}  \left[\alpha_{i }^{j}({\bf k})\textrm{e}^{i {\bf k} \cdot {\bf r}}\right]^* \sum_q g_q \left( b_{{\bf r},q} + b_{{\bf r},q}^\dagger \right),
\end{split}
\end{align}
where $j$ is a function of ${\bf r}$; it returns the color index at the ${\bf r}$ site. Using equation~\ref{fi0} we have, up to second order in the coupling amplitudes,
\begin{equation}
	i \hbar \dot{\Phi}_{i,{\bf k}} (t)= E_{i,{\bf k}} \Phi_{i,{\bf k}}(t) +\sum_{{\bf r} \in R_p}  \left[\alpha_{i }^{j}({\bf k}) \textrm{e}^{i {\bf k} \cdot {\bf r}} \right]^* V_{\bf r}^{\textrm{in}}(t).
\end{equation}
Now we use the Fourier transformation with respect to the time variable to write, still up to second order in the coupling amplitudes, 
\begin{equation} 
\Phi_{i,{\bf k}} (\omega)= \dfrac{1}{-\hbar \omega - E_{i,{\bf k}} + i 0^+} \sum_{{\bf r} \in R_p}  \left[\alpha_{i }^{j}({\bf k}) \textrm{e}^{i {\bf k} \cdot {\bf r}} \right]^* V_{\bf r}^{\textrm{in}}(\omega) ,
\end{equation} 
and 
\begin{equation} 
\Phi_{i,{\bf k}}^\dagger (\omega)= -\dfrac{1}{-\hbar \omega + E_{i,{\bf k}} + i 0^+} \sum_{{\bf r} \in R_p}  \alpha_{i }^{j}({\bf k}) \textrm{e}^{i {\bf k} \cdot {\bf r}} V_{\bf r}^{\textrm{in}}(\omega).
\end{equation} 
Note that $\Phi_{i,{\bf k}} (\omega) = \textrm{T.F.}\left( \Phi_{i,{\bf k}}\right)(\omega)$, T.F. denoting the Fourier transform and $\Phi_{i,{\bf k}}^\dagger (\omega) = \textrm{T.F.}\left( \Phi_{i,{\bf k}}^\dagger\right)(\omega)$ so $ \Phi_{i,{\bf k}} (\omega) \neq \left( \Phi_{i,{\bf k}}^\dagger (\omega) \right)^\dagger$. 

We introduce the $\beta_{i }^{j}({\bf k})$ coefficients such that
\begin{equation}
a_{j,{\bf k}}^\dagger \ket{0} = \sum_{i=1}^{N_C} \beta_{i }^{j}({\bf k}) \Phi_{i,{\bf k}}^\dagger \ket{0} .
\end{equation}
Then we have
\begin{equation}
x_{\bf r} = \dfrac{1}{ N} \sum_{\bf k} \sum_{i=1}^{N_C} \textrm{e}^{i {\bf k} \cdot {\bf r}} \left[\beta_{j({\bf r})}^i({\bf k})\right]^* \Phi_{i,{\bf k}}+ h.c. ,
\end{equation}
and we obtain, up to second order in the coupling amplitudes
\begin{equation} \label{eno}
x_{\bf r}(\omega) = \sum_{{\bf r}_0 \in R_p} \chi_{{\bf r},{\bf r}_0} V_{{\bf r}_0}^{\textrm{in}}(\omega),
\end{equation}
with
\begin{align}
\begin{split}
\chi_{{\bf r},{\bf r}_0} = \dfrac{1}{N}\sum_{\substack{ i=1 \\ {\bf k} }}^{N_C}  \left( \dfrac{C_{i,{\bf k},{\bf r},{\bf r}_0}^*}{-\hbar\omega - E_{i,{\bf k}} + i 0^+} - \dfrac{C_{i,{\bf k},{\bf r},{\bf r}_0}}{-\hbar\omega + E_{i,{\bf k}} + i 0^+} \right),
\end{split}
\end{align}
and
\begin{equation}
C_{i,{\bf k},{\bf r},{\bf r}_0}=\beta_{j({\bf r})}^i({\bf k}) \alpha^{j({\bf r}_0)}_{i}({\bf k}) \textrm{e}^{-i {\bf k} \cdot ({\bf r}-{\bf r}_0)}.
\end{equation}
Let us notice from the last equation that adding a probe with no input does not influence the response at the other probes (with or without input). This is because we restricted the response to first order in $g$. 

Finally, Eq.~\eqref{ano} gives, up to fourth order in the coupling amplitudes $\{g_q\}$, 
\begin{equation} \label{eq:hzfjzd1}
V_{{\bf r}}^{\textrm{out}}(\omega) =V_{{\bf r}}^{\textrm{in}}(\omega) +i J(\omega])\sum_{{\bf r}_0 \in R_p} \chi_{{\bf r},{\bf r}_0} V_{{\bf r}_0}^{\textrm{in}}(\omega).
\end{equation}

In the simple case ${\bf r}={\bf r}_0={\bf R}_0$ we consider in the main text, for simplicity, we define $\gamma_{j_0,{\bf k}}^{i}$ such that $\gamma_{j_0,{\bf k}}^{i}=C_{i,{\bf k},{\bf R}_0,{\bf R}_0} = \beta_{j_0 }^{i}({\bf k}) \alpha_{i }^{j_0}({\bf k})$. Moreover, for the Haldane system we consider the Bloch eigenvectors
\begin{equation} \label{eq:blockvec}
\ket{u_{i,{\bf k}}} = u_{i,{\bf k}}^\dagger \ket{0} = \left[\alpha_{i }^{1}({\bf k}) a_{1,{\bf k}}^\dagger + \alpha_{i }^{2}({\bf k}) a_{2,{\bf k}}^\dagger \right] \ket{0}, 
\end{equation}
which are defined through the components $\alpha_{i }^{1}({\bf k})$ and $\alpha_{i }^{2}({\bf k})$
\begin{equation} \label{eq:blockcompo}
\alpha_{i }^{2}({\bf k}) =  \dfrac{h_1({\bf k})^*}{h_2({\bf k})-h_0({\bf k})+E_{i,{\bf k}}} \alpha_{i }^{1}({\bf k}).
\end{equation}
We have introduced $E_{i,{\bf k}} = h_0({\bf k}) + (-1)^i \epsilon({\bf k})$, $\epsilon({\bf k}) = \sqrt{|h_1({\bf k})|^2+h_2({\bf k})^2}$ and the coefficients $\beta_{i }^{j}({\bf k})$ satisfy 
\begin{equation} \label{eq:blockreverse}
a_{j,\boldsymbol{k}}^\dagger \ket{0} = \sum_{i=1}^2 \beta_{i }^{j}({\bf k}) \Phi_{i,\boldsymbol{k}}^\dagger \ket{0}, \, j=\{A,B\}.
\end{equation}
Using Eqs.~\eqref{eq:blockvec}, \eqref{eq:blockcompo} and \eqref{eq:blockreverse}, we obtain
\begin{equation}
\gamma_{j_0,{\bf k}}^{i}  = \dfrac{1}{2}+\dfrac{(-1)^{j_0+i+1} h_2({\bf k})}{2 \epsilon({\bf k})}.
\end{equation}